# MODELING BUSINESS

**Audris Kalnins, Valdis Vitolins**

*University of Latvia, IMCS, 29 Raina blvd, LV-1459, Riga, Latvia*

*audris@cclu.lv, valdis_vitolins@exigengroup.lv*

In this article business concepts are studied using a metamodel-based approach. The Notation Independent Business concepts metamodel is introduced. The approach offers a mapping between different business modeling notations which could be used for bridging BM tools and boosting the MDA approach

## 1 Modeling Problems

An unambiguous and formal method how to describe some problem has been demanded always. A model of the problem is essential for [1]:

- investigation and analysis,
- unambiguous information exchange between people and computer systems,
- development of a practical solution, using the problem model.

Additionally a model must support measurement and comparison (both for simulation and in real environment), and should support version and change management during the model life cycle.

To ensure now a successful business, an enterprise must integrate its information systems with partners and make the data exchange available. Today not only a simple data exchange but also more complicated *service exchange* is already used (e.g. web services and e-business). However, the demands are even higher – the business should support not only a simple service at a narrow area, but also *any service at any place* (e.g. e-government). In order to build such interoperable process definitions a notation independent process modeling is of highest value.

Currently many *business modeling* (BM) languages exist and even more tools, which support them. The most popular ones are ARIS, IDEF3, UML with its Activity Diagram (AD), GRADE BM (BM) [2]. What refers to UML, its newest (in-progress) version 2.0 [3] is referenced in this paper. However, these languages mainly support modeling of the business processes themselves, but say nothing or very little about the business process environment and semantics. They don't answer - why this process is used, how it is linked to other processes and enterprise demands, how we can control and improve it. Certainly, some of these languages can be used as a framework for modeling the above mentioned higher-level concepts in an indirect way.

The question is – what are general concepts to which we should pay attention to describe all aspects of an enterprise, and which is the best way? Our approach to this question is *metamodel* based. We would like to build a metamodel, which would be:

- simple and natural enough to be understood also by non-IT people
- comprehensive enough cover all business aspects which are more or less related to business processes
- detailed enough to serve the goals of system description, analysis and design, as far as business processes are concerned
- serve as a common basis for the most typical BM notations and would enable a more or less automated concept mapping from one notation (modeling language) to another.

Thus our goal would not be to invent a new BM notation, but to provide more understanding to the existing ones and means to harmonize and extend them. Certainly, it is impossible to cover literally all aspects of the business (see a little bit failed attempt of this in [4]), therefore we will concentrate on what is "around the processes". It should be noted that such "platform independent modeling" is completely in line with the MDA (Model Driven Architecture) approach, extending the platform independent development to process modeling area.

## 2 Business Metamodels

Business modeling is already a widespread term, however we wish to say that only some *views* of business have been investigated. At this time unambiguous and well-defined models exist only for several narrow business areas, but wide and comprehensive models are very informal and generic. For example, a business process as a sequence of activities is well defined in several notations, but practically nothing is said about its environment.

J. Zachman is one of the first, who has investigated enterprise level business aspects. His methodology[5], developed in 1987. and named *Zachman Framework*, has gained popularity and has been implemented in several modeling tools, including *System Architect* [6] by Popkin Software. In this framework, several high-level business concepts are introduced - goals, strategy and business plan. These terms are used as a context for concepts, included in model diagrams. Unfortunately, this framework is too informal and doesn't show concept relationships, no metamodeling approach has been used there.

As e-business spreads more widely, many models have been developed from this prospective. Quite an accurate view was introduced by Andreas Dietzsch [7]. His approach is described by a metamodel – a class diagram (Fig. 1) therefore it is quite unambiguous. The model shows the business goals and strategy like Zachman framework, but it also introduces input-output chain of an enterprise. Some of the model concepts (supplier, input, business process, output, customer) have a good correlation with the ISO/DIS standard[8]. This model introduces several types of enterprise processes (management, support, performing).

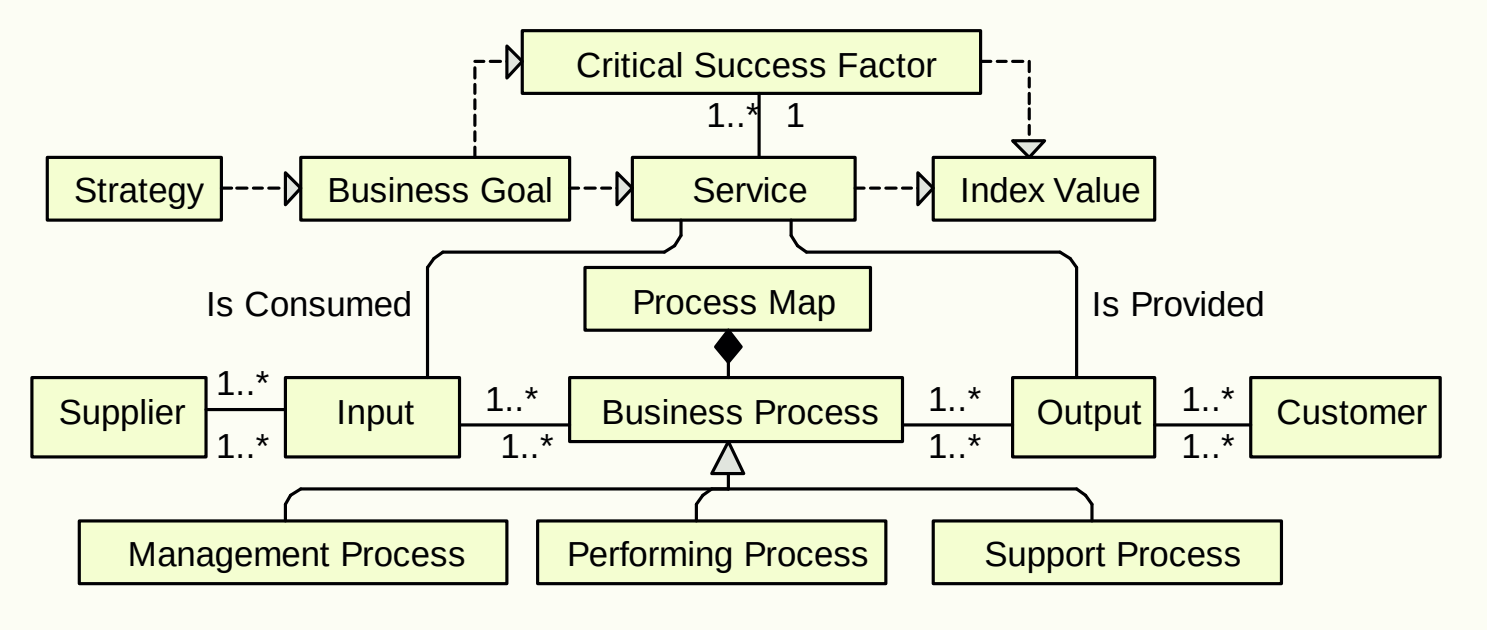

**Fig.1 Business metamodel by A. Dietzsch**

Another view of business is developed within the *COMBINE* project [9]. In this project the main concepts of business were investigated in relation to distributed systems (due to space limitations, some lower level classes are not displayed in Fig. 2):

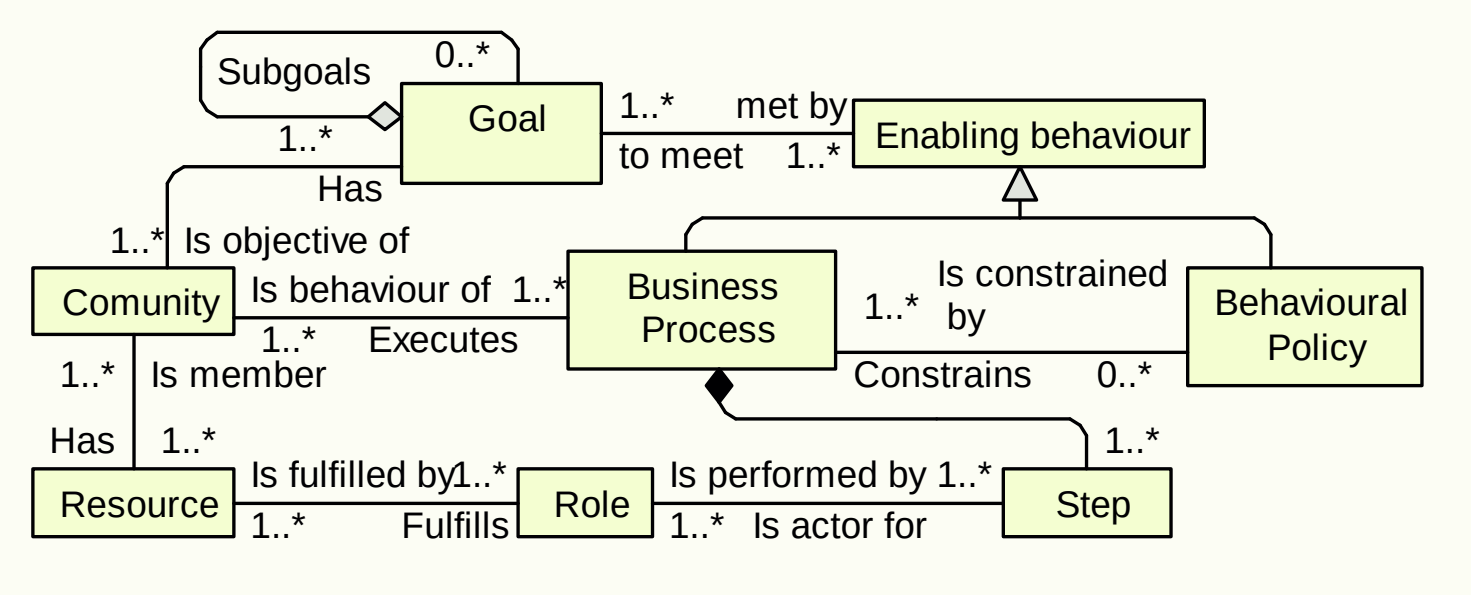

**Fig.2 Business metamodel from Combine project**

In this model resources and their relation to business processes are highlighted quite well. By the way, in this model the *Community* is what others call an *Enterprise*.

As you can see, the above-mentioned metamodels are quite different, yet a more dissimilar view on business concepts is described in [10], where business systems are explored in terms of e-business requirements. One more view on Business Processes is presented by Business Process Management Initiative (BPMI) [11], where a business is modeled in terms of its interfaces and collaboration. Sure, we could list more models - detailed or generic, but the ones mentioned here are enough to show how different the views can be on the same area.

## 3 Main business concepts and their relations

When several sources are combined into one model, the main problem is that similar concepts are named very differently. Therefore, the first step is the analysis of semantics of concepts, and harmonization to one general concept (e.g., *Community* and *Enterprise*). To support a general business pattern, we choose the *Value Added Chain* approach - a business of an enterprise is described in terms of input (the enterprise is supplied with raw materials or "raw" service), process (add some value to this) and output (deliver it to customer as a value added service or a processed material)[12].

To make our metamodel as small as possible we paid attention only to these aspects, which play some role in business process automation. Such things as *competitors* or *reputation* that play big role in business, but are not directly related to business processes, are not included. However, we included such "intangible" thing as *Knowledge*, addressing that part of the enterprise knowledge, which is stored as data, especially in knowledge management systems becoming so popular.

Also, we tried to exclude as many abstract classes from the model as possible, because they damage the model understandability. For sample, we included real classes *Requirement* and *Conformity*, but we didn't include *Quality* (a conformity to determined requirements). *Quality* is shown only as a two-way dependency between these classes. We made an exclusion regarding to the abstract class *Performer*. It was introduced due to wish to provide a more general relation to the *Task* concept.

Fig. 3 presents our current proposal for the business metamodel.

In the presented model, *Business Goal* means anything (tasks, goals, plans), that is strategic and for a long term. *Business Goal* [8] is split into two parts – *Strategy*, witch usually is an unstructured information and *Index Value* – a measurable magnitude that is used to measure the progress in strategy implementation. *Resource* and its relation to *Enterprise*, *Input*, *Output* and *Business Process* is taken from the *Value Added Chain* approach. The *Measurement-Refinement* and *Determination-Satisfaction* chains conform to ISO. The important terms *Business Process* and *Performer* are described in more detail in a separate metamodel fragment - Fig. 4.

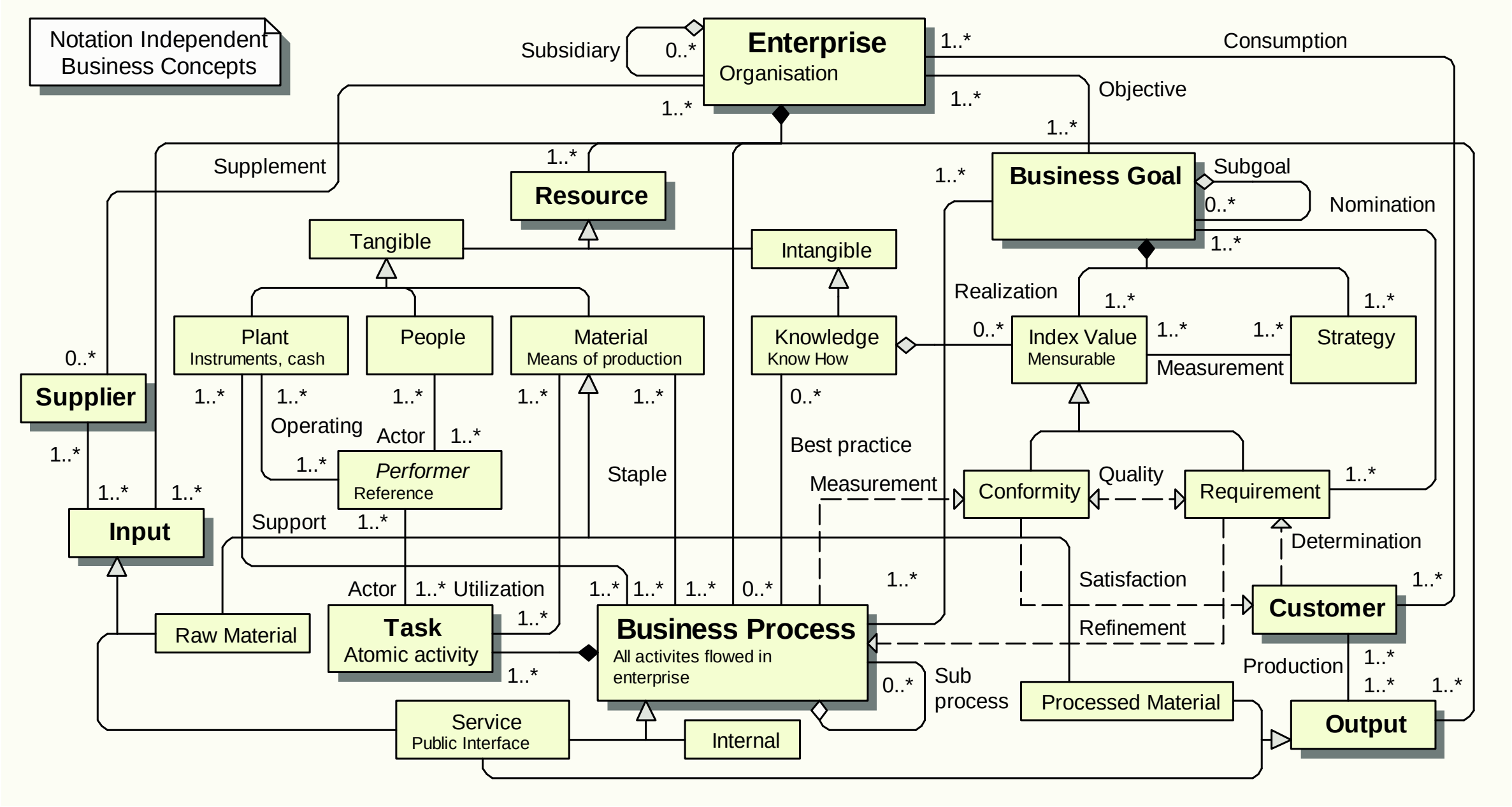

**Fig. 3. Business metamodel – the top classes.**

As mentioned before, the *Performer* is an abstract class (reference), which points to real resources (explicitly to *Resource* or implicitly through other references - *Organizational Unit*, *Role* (a role in a project) or *Qualification*. *Task* is an atomic business process unit, which actually describes some step or function and is done by a *Performer*. In some notations it is called *Action* (UML 2.0 Activity diagram), *Task* (GRADE BM, BPMN), *Unit of Behavior* (IDEF3). *Transition* determines the sequence in which several *Tasks* are executed. *Pass* is a normal transition from task to task. Due to differences in notations (GRADE PM and BPMN incorporate task triggering and output branching in the task itself while UML or IDEF3 does not), separate subclasses – *Incoming* and *Outgoing* transitions must be introduced. These classes will be explained in detail in the following section. *Control* is singled out as a separate abstract class. It includes *Decision* (start of control branching), *Fork* (start of concurrent threads), flow unification (*Merge* and *Join* – for concurrent threads) and process start and stop points.

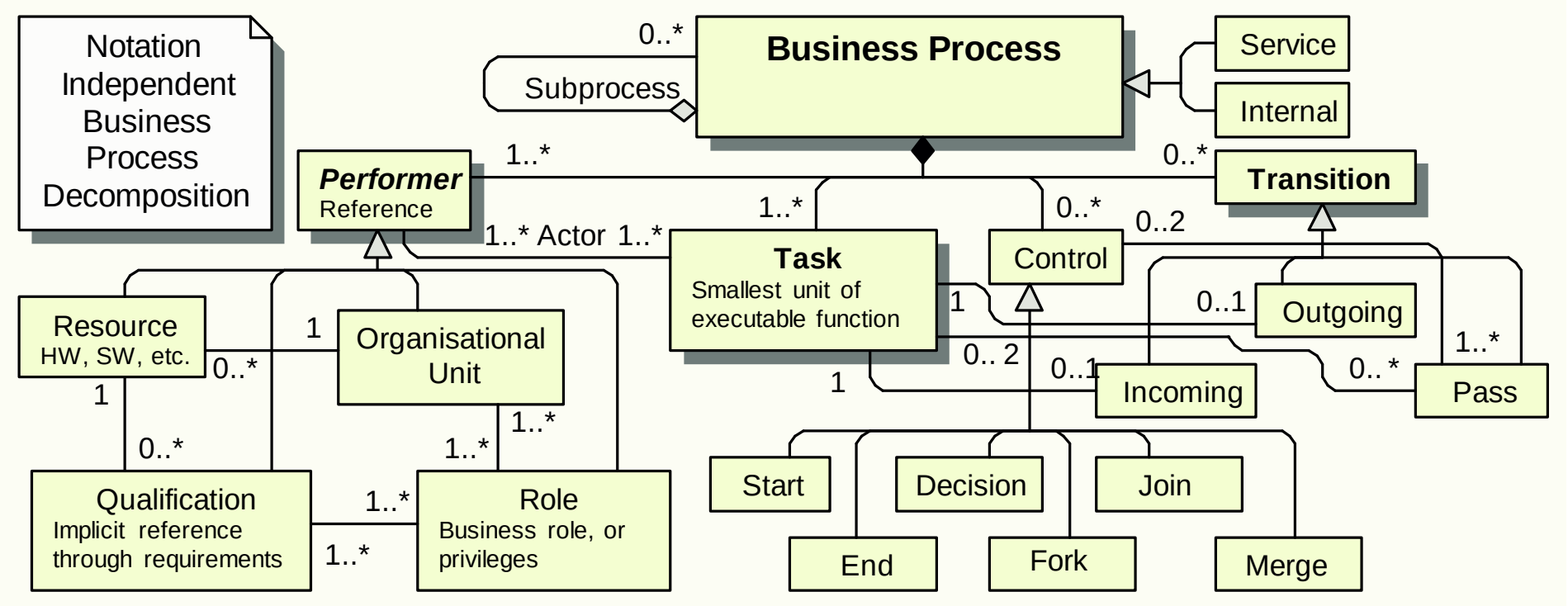

**Fig. 4 Business metamodel – details of Business Process**

## 4 Concept mapping in different notations

This section deals with the problem how the metamodel of business concepts can be used to define a relation between different business modeling notations. Namely, a *mapping* between such notations will be defined. The approach will be illustrated by defining a mapping between GRADE BM and UML AD notations (actually a fragment of it), but it could be extended to other notations as well.

It should be noted that there is no special support for defining a mapping between two class diagrams in UML – a fact acknowledged also in the MDA area. However, when the mapping is at "class level" (i.e., an instance of a class in one diagram is always mapped to an instance of the corresponding class (or instances of several classes) in the other diagram) the mapping can be denoted via associations between corresponding classes (augmented when necessary by constraints such as XOR). Namely this is the case in the paper.

Another issue is the graphical presentation of concepts in business modeling by diagram elements. For example, the generic solution of this problem in UML 2.0 is positioned outside the main UML metamodel. Fortunately, for business processes (Activity diagrams in UML 2.0) there exists already a “logical diagram concept” (*Activity*) in the metamodel, in the result the “UML-internal” mapping from domain concepts to their graphical presentation actually is one-one. The situation is similar in most BM notations, including GRADE.
Therefore it is natural in the BM area to define a mapping at the domain layer. Namely, we take the “native” metamodels of several BM notations (in our case, GRADE BM and UML AD) and define a pair of mappings from their concepts to the concepts in our notation independent metamodel. They are shown as associations between the corresponding classes of the metamodels. A concept from a particular notation can be mapped to one or more concepts in the independent domain (and vice versa). In the result, a derived mapping (which can be quite complicated) is obtained between the graphical elements of both notations – namely this mapping is of interest for the users of BM notations. Fig . 5 summarizes this approach at meta-metamodel level, showing the possible primary and derived associations between two notations A and B and the independent metamodel.

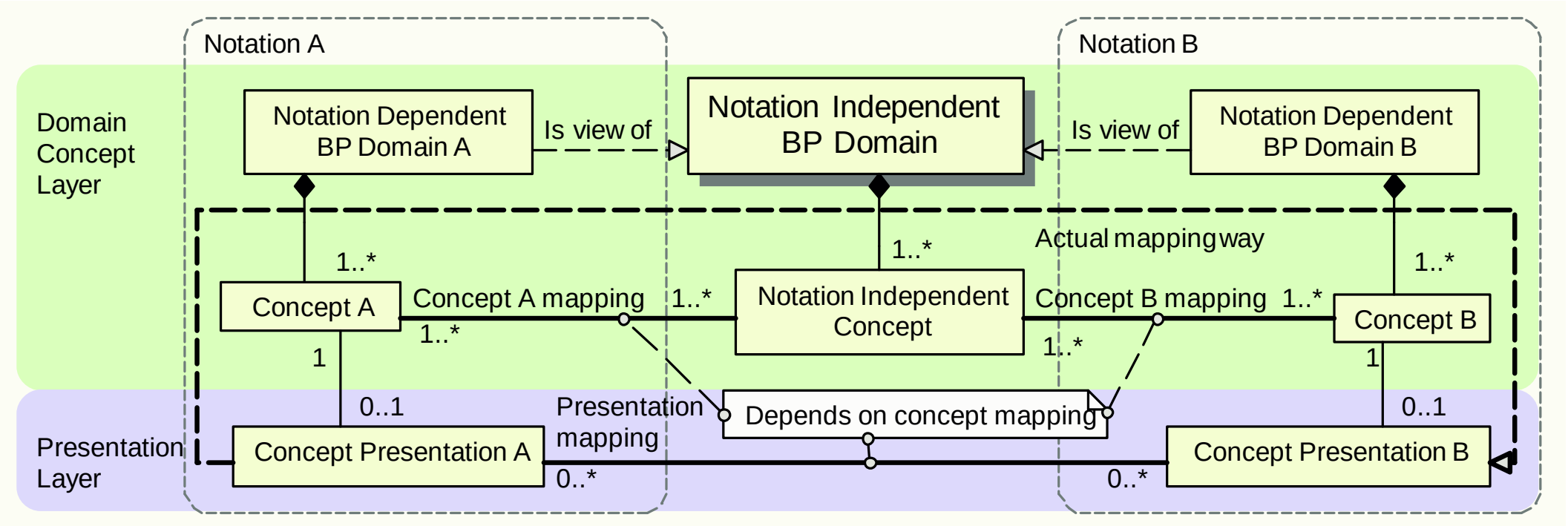

**Fig.5 The scheme of metamodel mapping**

It should be noted that another approach to notation mapping could be used – namely that in the Exigen Business Modeler [2]. There one common domain metamodel (the notation independent one) is mapped to several notation dependent graphical layer metamodel fragments. This approach is more efficient for tool building, but seems to be more limited in power – though GRADE BM and UML AD have been mapped to each other, adding more notations would require significant changes in the “fine-tuned” domain metamodel.
Unfortunately, the size of the paper does not permit to describe our mappings in detail. Instead, they will be explained on an example. A simple business process is displayed in Fig. 6 both as a GRADE BM and UML Activity diagram (the letters beside the lines are not part of the notation but annotate the mapping example).

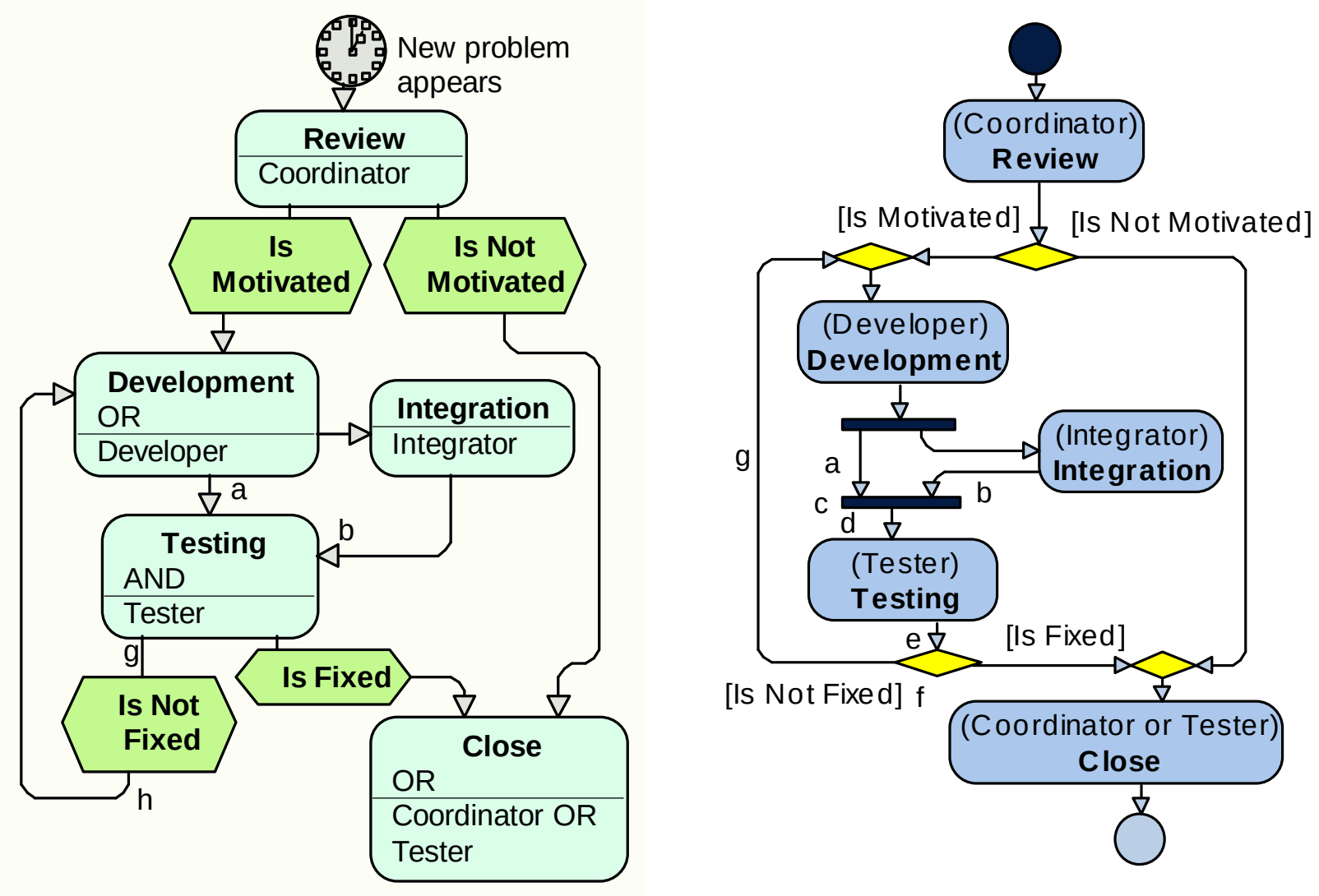

**Fig.6 A business process as GRADE BM and UML AD**

The next Fig. 7 shows part of the defined mapping between the GRADE metamodel classes and the independent domain (the top left area) and the independent domain and a fragment of UML 2.0 metamodel (top right). The (not shown) association multiplicities are 0..1 or 1, some XOR constraints are also shown. Note that the GRADE *Task* is a very inclusive concept – it includes a *triggering condition* attribute, and according to the values of that attribute (*OR*, *AND*, *none*) a task maps to either *Merge* or *Join* classes (or none of them) in the independent domain. If either of these classes is present the *Incoming* transition is also included in the mapping. A similar situation is for the task exit. It should be noted that the described conditional mapping can be described formally using OCL.

The lower area in Fig. 7 presents the mapping of some of the class instances according to the example in Fig. 6. Actually there should be instances also in the middle column, but they are omitted to simplify the drawing, since in this fragment the mapping between the notation independent classes and UML is one-one. The (not shown) mapping between the notation domains and their presentations is also one-one.

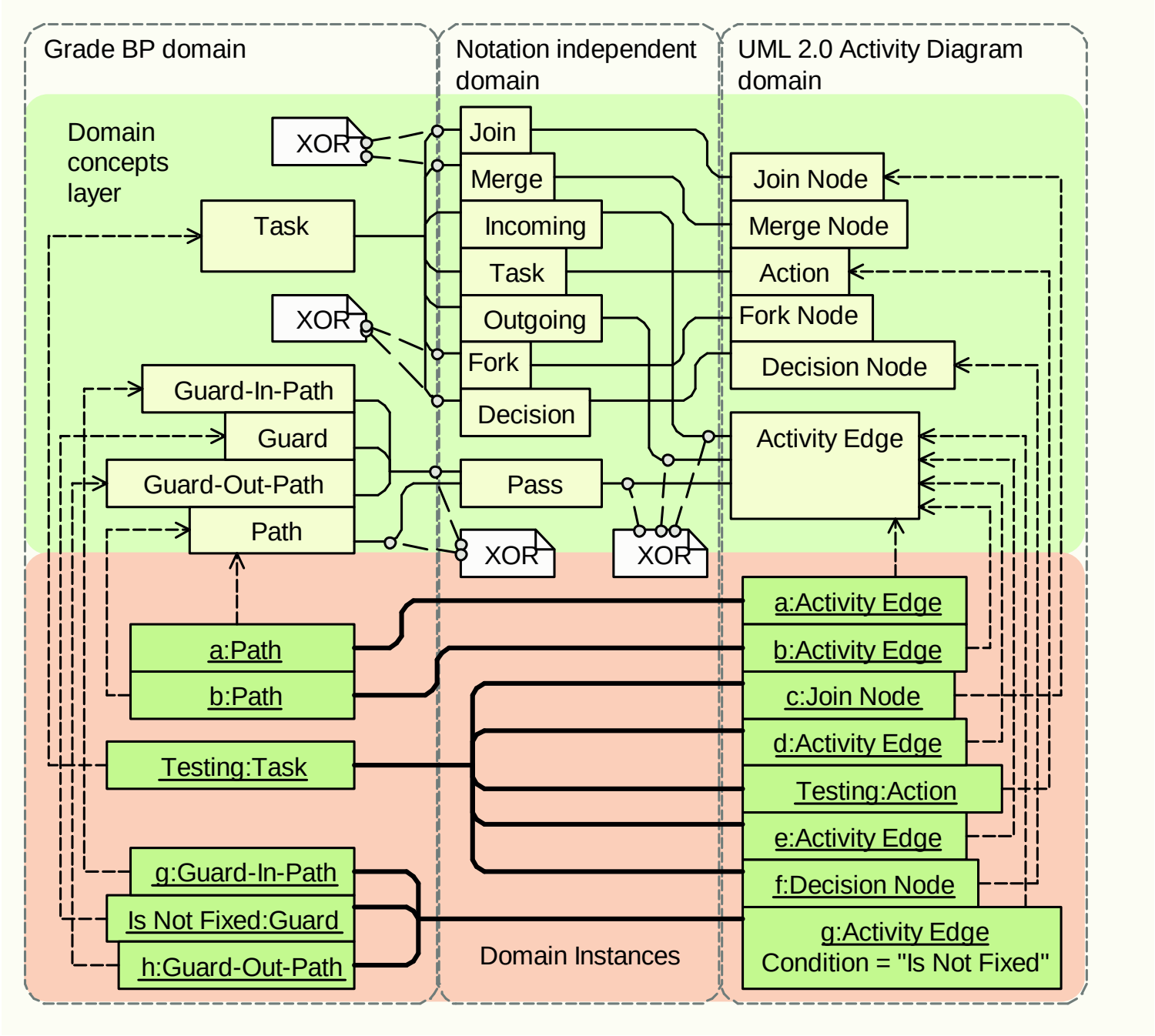

**Fig.7 Mapping definition and example**

## 5 Conclusion

The presented Notation independent business model and the mappings between two BM notations based on this metamodel is the first step in the selected direction. The results obtained so far seem to be promising for building a comprehensive business modeling framework covering most of the known BM notations and extending them. The practical aspect of this approach would be a possibility to build a notation independent repository for several simultaneously accessing BM tools using different BM notations. Each of the tools internally would "live" upon its native domain metamodel. Additionally this could boost the MDA approach by providing unambiguous model translations between different modeling notations.